\documentstyle[12pt,preprint,eqsecnum,aps]{revtex}
\tighten
\begin{document}
\preprint{MKPH-T-95-09}
\draft
\title{Compton scattering from a pion and off--shell effects\thanks{
Invited talk at the Conference on {\em Perspectives in Nuclear Physics at
Intermediate Energies}, \mbox{Trieste}, Italy, 8 -- 12 May 1995}
}
\author{S.\ Scherer\thanks{Supported by Deutsche Forschungsgemeinschaft}}
\address{Institut f\"ur Kernphysik, Johannes Gutenberg--Universit\"at,
D--55099 Mainz, Germany}
\maketitle
\begin{abstract}
   We discuss the most general form of the electromagnetic vertex of an
off--shell pion.
   The framework of chiral perturbation theory is used to illustrate
the representation dependence of the electromagnetic three--point Green's
function.
   For that purpose we discuss the concept of field transformations which,
in comparison with the standard Gasser and Leutwyler Lagrangian, generate
equivalent Lagrangians containing additional terms at order $p^4$
proportional to the lowest--order equation of motion.
   We consider Compton scattering from a pion to show that calculations
involving different off--shell effects in the s-- and u--channel pole diagrams
may nevertheless lead to the same on--shell Compton scattering amplitude.
   This is a result of the equivalence theorem which states that two
equivalent Lagrangians predict the same S--matrix elements even though they
may generate a different off--shell behaviour of Green's functions.
   We conclude that off--shell effects are not only model dependent but
also representation dependent.
\end{abstract}
\section{Introduction}
   A covariant description of almost any electromagnetic process involving
pions will, in principle, require the off--shell electromagnetic vertex
of a pion \cite{Nishijima} which, in general, is expected to be more
complicated than the free vertex.
   Let us consider a few examples in order to illustrate this point.
   The t--channel pole diagram of pion photo-- and electroproduction off a
nucleon contains a virtual pion which interacts with the photon and
then emerges as an on--shell pion.
   Similarly, pion--nucleon bremsstrahlung will involve four pole diagrams
where the photon is radiated off the external pion lines.
   One pion line is off shell, the other on shell.
   A situation where both pion lines are off mass shell is encountered
in meson exchange current contributions to electron scattering from nuclei.
   Clearly, the above and many other examples suggest that a systematic study
of off--shell effects in electromagnetic interactions is important.

   The purpose of this talk is to deal with the following three issues:
\begin{enumerate}
\item What is the most general form of the electromagnetic vertex of an
off--shell pion?
\item How does this vertex show up in the calculation of an invariant
amplitude such as, e.g., the Compton scattering amplitude?
\item Is it, in principle, possible to obtain information about off--shell
contributions from experimental data?
\end{enumerate}
   In order to answer these questions we shall make use of the framework of
chiral perturbation theory (ChPT) for mesons \cite{Weinberg,Gasser1,Gasser2}.
   Use of a specific theory in the calculation of a simple process, namely
$\gamma\pi^+\to\gamma\pi^+$, will serve us to understand the nature of
off--shell contributions.
   The interpretation of the result, however, is general and will not depend
on the fact that we have performed a calculation at $O(p^4)$ in the
momentum expansion.

\section{Electromagnetic vertex of an off--shell pion}
   In this section we shall formally introduce the concept of {\em form
functions} by parameterizing the electromagnetic three--point Green's function
of a pion.
   We deliberately distinguish between {\em form factors} and
{\em form functions}, since form factors correspond to observables, which is
not necessarily true for form functions.
   Then a discussion of some general properties of these form functions,
resulting from the application of symmetry principles, follows.

   Let us consider the momentum--space Green's function of two unrenormalized
pion field operators $\pi^+(y)$ and $\pi^-(z)$ coupled to the electromagnetic
current operator $J^\mu(x)$,
\footnote{$\pi^{+/-}(x)$ destroys a $\pi^{+/-}$ or creates a $\pi^{-/+}$.}
\begin{eqnarray}
\label{gmu}
\lefteqn{(2\pi)^4 \delta^4(p_f-p_i-q) G^{\mu}(p_f,p_i)=}\nonumber\\
&&\int d^4x\, d^4y\, d^4z\, e^{-i(q \cdot x - p_f \cdot y  + p_i \cdot z )}
<\!\! 0|T\left(J^{\mu}(x) \pi^+(y) \pi^-(z)\right)|0\!\!>,
\end{eqnarray}
where $p_i$ and $p_f$ are the four--momenta corresponding to the pion lines
entering and leaving the vertex, respectively, and $q=p_f-p_i$ is the momentum
transfer at the vertex.
   The renormalized three--point Green's function $G^\mu_R$ is defined as
\cite{Cheng}
\begin{equation}
\label{gmur}
G^{\mu}_R(p_f,p_i) = Z^{-1}_{\phi} Z^{-1}_J G^{\mu}(p_f,p_i),
\end{equation}
where $Z_{\phi}$ and $Z_J$ are renormalization
constants.\footnote{In fact, $Z_J=1$ due to gauge invariance \cite{Cheng}.}
   The irreducible, renormalized three--point Green's function is then given
by
\begin{equation}
\label{gammamuirr}
\Gamma^{\mu,irr}_R(p_f,p_i) =
(i \Delta_R(p_f))^{-1} G^{\mu}_R(p_f,p_i)(i\Delta_R(p_i))^{-1},
\end{equation}
where $\Delta_R(p)$ is the full, renormalized propagator.
   From a perturbative point of view, $\Gamma^{\mu,irr}_R$ is made up of those
Feynman diagrams which cannot be disconnected by cutting any one single
internal line.

   We shall now discuss some model--independent properties of
$\Gamma^{\mu,irr}_R(p_f,p_i)$ which follow from general symmetry
considerations, such as Lorentz covariance, time--reversal and
charge--conjugation symmetry, as well as gauge invariance.
\begin{enumerate}
\item Imposing Lorentz covariance, the most general parameterization of
$\Gamma^{\mu,irr}_R$ can be written in terms of two independent four--momenta,
$P^\mu=p_f^\mu+p_i^\mu$ and $q^\mu=p_f^\mu-p_i^\mu$, respectively,
multiplied by Lorentz--scalar form functions $F$ and $G$ depending on
three scalars, e.g., $q^2, p_i^2, p_f^2$,
\begin{equation}
\label{par}
\Gamma^{\mu,irr}_R(p_f,p_i) =
(p_f+p_i)^{\mu} F(q^2,p_f^2,p_i^2) +
(p_f-p_i)^{\mu} G(q^2,p_f^2,p_i^2).
\end{equation}
\item Imposing time--reversal symmetry one finds the following properties
of the form functions,
\begin{equation}
\label{trs}
F(q^2,p_f^2,p_i^2)=F(q^2,p_i^2,p_f^2), \quad
G(q^2,p_f^2,p_i^2)=-G(q^2,p_i^2,p_f^2).
\end{equation}
   In particular, from Eq.\ (\ref{trs}) we conclude that $G(q^2,M^2,M^2)=0$.
   This, of course, corresponds to the well--known fact that a spin--0
particle has only one electromagnetic form factor, $F(q^2)$.
\item Using the charge--conjugation properties $J^\mu\to-J^\mu$ and
$\pi^+\leftrightarrow\pi^-$, it is straightforward to see that form
functions of particles are just the negative of form functions of
antiparticles.
   In particular, the $\pi^0$ does not have any electromagnetic form functions
even off shell, since it is its own antiparticle.
\item The equal--time commutation relations of the electromagnetic
charge--density operator with the pion field operators,
\begin{eqnarray}
\label{comrel}
[J^0(x),\pi^-(y)] \delta (x^0-y^0) & = & \delta^4(x-y) \pi^-(y), \nonumber \\
{[}J^0(x),\pi^+(y)]  \delta (x^0-y^0) & = & -\delta^4(x-y) \pi^+(y),
\end{eqnarray}
and current conservation at the operator level, $\partial_\mu J^\mu(x)$,
are the basic ingredients for obtaining Ward--Takahashi identities
\cite{Ward,Takahashi} relating Green's functions which differ by one
insertion of the electromagnetic current operator.
   In particular, one obtains for the electromagnetic vertex
\begin{equation}
\label{wti}
q_{\mu} \Gamma^{\mu,irr}_R(p_f,p_i) =
\Delta_R^{-1}(p_f)-\Delta_R^{-1}(p_i).
\end{equation}
   Inserting the parameterization of the irreducible vertex, Eq.\ (\ref{par}),
into the Ward--Takahashi identity, Eq.\ (\ref{wti}), the form functions
$F$ and $G$ are constrained to satisfy
\begin{equation}
\label{constraint}
(p_f^2-p_i^2) F(q^2,p_f^2,p_i^2)+q^2 G(q^2,p_f^2,p_i^2)
= \Delta^{-1}_R(p_f)-\Delta^{-1}_R(p_i).
\end{equation}
   From Eq.\ (\ref{constraint}) it can be shown that, given a consistent
calculation of $F$, the propagator of the particle, $\Delta_R$, as well as
the form function $G$ are completely determined (see Appendix A of
Ref.\ \cite{Rudy} for details).
   The Ward--Takahashi identity thus provides an important consistency check
for microscopic calculations.
\end{enumerate}

\section{Chiral perturbation theory for mesons}
   In this section we shall give a short introduction to those aspects of
chiral perturbation theory \cite{Weinberg,Gasser1,Gasser2} which are relevant
for our discussion of off--shell Green's functions.
   In particular, the concept of field transformations is discussed in
quite some detail since it turns out to be important for the interpretation
of form functions.

   Chiral perturbation theory provides a systematic framework for describing
interactions between the members of the low--energy pseudoscalar octet
$(\pi,K,\eta)$ which are regarded as the Goldstone bosons of spontaneous
symmetry breaking in $QCD$ from $SU(3)_L\times SU(3)_R$ to $SU(3)_V$.
   The effective Lagrangian of ChPT is organized as a sum of terms with an
increasing number of covariant derivatives and quark mass terms,
\begin{equation}
\label{lag}
{\cal L}={\cal L}_2 + {\cal L}_4 + \dots,
\end{equation}
where the subscripts refer to the order in the momentum expansion.
   Covariant derivatives and quark mass terms are counted as $O(p)$ and
$O(p^2)$, respectively, in the power counting scheme.
   A systematic classification of Feynman diagrams in the framework of an
effective Lagrangian is made possible by Weinberg's power counting scheme
\cite{Weinberg} which, loosely speaking,\footnote{We do not address
the question of renormalization.} can be interpreted as follows.
   Given a general diagram calculated, e.g., with the Lagrangian of
Eq.\ (\ref{lag}), a rescaling of all external momenta $p\to tp$ and of masses
$M^2\to t^2 M^2$ leads to the following behaviour of the corresponding
invariant amplitude\footnote{Polarization vectors also have to be rescaled,
$\epsilon\to t\epsilon$.}
\begin{displaymath}
{\cal M}(tp,t^2 M^2)=t^D {\cal M}(p,M^2).
\end{displaymath}
   The scaling power $D$ is determined by
\begin{equation}
\label{dimd}
D=2+\sum_{n=2,4,\dots}(n-2)N_n+2N_L,
\end{equation}
where $N_{n}$ refers to the number of vertices for which the number of
covariant derivatives $n_1$ and twice the number of quark mass terms or field
strength tensors $n_2$ adds up to $n$, and where $N_L$ denotes the number of
loops.
   For small enough external momenta and masses which, of course, can
only be controlled theoretically, diagrams with large dimensions $D$ are
expected to be less important.
   According to Eq.\ (\ref{dimd}), at $O(p^2)$ one has to consider
tree--level diagrams constructed entirely with vertices from ${\cal L}_2$,
at $O(p^4)$ one has to take account of tree--level diagrams with one
vertex from $O(p^4)$ and an arbitrary number of vertices from $O(p^2)$
or one--loop diagrams with vertices from ${\cal L}_2$.

   The lowest--order Lagrangian is given by \cite{Gasser2}
\begin{equation}
\label{l2}
{\cal L}_2 = \frac{F_0^2}{4} Tr \Big ( D_{\mu} U (D^{\mu}U)^{\dagger}
+\chi U^{\dagger}+ U \chi^{\dagger} \Big ),\quad
U(x)=\exp\left( i\frac{\phi(x)}{F_0} \right ),
\end{equation}
where
\begin{equation}
\label{phi}
\phi(x)=\left (
\begin{array}{ccc}
\pi^0+\frac{1}{\sqrt{3}}\eta & \sqrt{2} \pi^+ & \sqrt{2} K^+ \\
\sqrt{2} \pi^- & -\pi^0+\frac{1}{\sqrt{3}}\eta & \sqrt{2} K^0 \\
\sqrt{2} K^- & \sqrt{2} \bar{K}^0 & -\frac{2}{\sqrt{3}}\eta
\end{array}
\right ).
\end{equation}
The quark mass matrix which in $QCD$ generates an explicit breaking of chiral
symmetry is contained in $\chi=2 B_0\, diag(m_u,m_d,m_s)$.
$B_0$ is related to the quark condensate $<\!\!\bar{q}q\!\!>$,
$F_0\approx 93$ MeV denotes the pion--decay constant in the chiral
limit.
   The covariant derivative $D_\mu U = \partial_\mu U +ie A_\mu [Q,U]$,
where $Q=diag(2/3,-1/3,-1/3)$ is the quark--charge matrix, $e>0$,
generates a coupling to the electromagnetic field $A_\mu$.
   Finally, the equation of motion (EOM) obtained from ${\cal L}_2$ reads
\begin{equation}
\label{eom2}
{\cal O}^{(2)}_{EOM}(U)=D^2 U U^\dagger - U (D^2 U)^\dagger
-\chi U^\dagger +U \chi^\dagger
+\frac{1}{3}Tr\left(\chi U^\dagger-U\chi^\dagger\right)=0.
\end{equation}
The most general structure of ${\cal L}_4$ was first written down by Gasser and
Leutwyler (see Eq.\ (6.16) of Ref.\ \cite{Gasser2}),
\begin{equation}
\label{l4}
{\cal L}_4=L_1\left(Tr(D_\mu U (D^\mu U)^\dagger)\right)^2 + \dots,
\end{equation}
and introduces 10 physically relevant low--energy coupling constants $L_i$.

   Before turning to the Compton scattering process we still have to discuss
the important concept of field transformations
\cite{Kamefuchi,Coleman,Scherer}.
   For that purpose we introduce a field redefinition,
\begin{equation}
\label{up}
U'=exp(iS)U=U+iSU+\dots,
\end{equation}
and demand the same properties of the new fields $U'$ as of $U$.
   For example, since both, $U$ and $U'$, are $SU(3)$ matrices we conclude
that $S=S^\dagger$ and $Tr(S)=0$.
   Furthermore, imposing the constraints of chiral symmetry, charge conjugation
and parity it can be shown that two generators exist at $O(p^2)$
(see Ref.\ \cite{Scherer} for details).

   What is the consequence of working with $U'$ instead of $U$?
   Inserting $U'$ into ${\cal L}$ of Eq.\ (\ref{lag}) results in
\begin{equation}
\label{lagup}
{\cal L}(U)\rightarrow{\cal L}(U')={\cal L}_2(U)+\Delta {\cal L}_2(U)
+{\cal L}_4(U) + O(p^6),
\end{equation}
where
\begin{equation}
\label{dlag2}
\Delta {\cal L}_2(U)= \mbox{tot.div.} +\underbrace{\frac{F^2_0}{4}
Tr(iS {\cal O}_{EOM}^{(2)})}_{O(p^4)} + \underbrace{O(S^2)}_{O(p^6)}.
\end{equation}
   The total divergence has no dynamical significance and can thus be dropped.
   The second term of Eq.\ (\ref{dlag2}) is of $O(p^4)$ and leads to a
``modification'' of ${\cal L}_4$ \cite{Rudy,Scherer},
\begin{equation}
\label{loffshell}
{\cal L}^{off-shell}_4=\beta_1 Tr({\cal O}_{EOM}^{(2)}
{\cal O}^{(2)\dagger}_{EOM})
+\beta_2Tr((\chi U^\dagger-U\chi^\dagger){\cal O}_{EOM}^{(2)}).
\end{equation}
   By a simple redefinition  of the field variables one generates an infinite
set of ``new'' Lagrangians depending on two parameters $\beta_1$ and $\beta_2$.
   That all these Lagrangians describe the same physics will be illustrated
in the next section.

\section{The Compton scattering amplitude}
   In this section we shall consider the invariant amplitude for the process
$\gamma (\epsilon,k)+\pi^+(p_i)\rightarrow \gamma'(\epsilon',k')+\pi^+(p_f)$
at $O(p^4)$ in the framework of the effective Lagrangians of Eqs.\ (\ref{l2}),
(\ref{l4}) and (\ref{loffshell}).
   The first discussion of this process in standard ChPT ($\beta_1=\beta_2=0$)
can be found in Refs.\ \cite{Bijnens,Donoghue}.
   Our main interest here is to investigate the influence of the additional
terms of Eq.\ (\ref{loffshell}) on a) the form functions of the pion and
b) the total Compton scattering amplitude.
   In particular, we want to find out whether the empirical Compton scattering
amplitude can be used to obtain information about the form functions
of the pion.

    The most general, irreducible, renormalized three--point Green's
function (see Eq.\ (\ref{par})) at $O(p^4)$, compatible with the
constraints imposed by approximate chiral symmetry, was derived in
Ref.\ \cite{Rudy}.
    For positively charged pions and for real photons ($q^2=0, q=p_f-p_i$)
it has the simple form
\begin{equation}
\label{emv}
\Gamma^{\mu,irr}_R(p_f,p_i)=(p_f+p_i)^{\mu}
\left(1+16\beta_1\frac{p_f^2+p_i^2-2M^2_{\pi}}{F^2_{\pi}}\right),
\end{equation}
and the corresponding renormalized propagator is given by
\begin{equation}
\label{propagator}
i\Delta_R(p)=\frac{i}{p^2-M^2_{\pi}
+\frac{16 \beta_1}{F^2_{\pi}}(p^2-M^2_{\pi})^2+i\epsilon}.
\end{equation}
   Note that Eqs.\ (\ref{emv}) and (\ref{propagator}) satisfy the
Ward--Takahashi identity, Eq. (\ref{wti}).
   Clearly, the parameter $\beta_1$ is related to the deviation from a
``pointlike'' vertex, once one of the pion legs is off shell.
   The electromagnetic vertex and the propagator are both independent of the
parameter $\beta_2$ of Eq.\ (\ref{loffshell}).
   Eqs.\ (\ref{emv}) and (\ref{propagator}) have to be compared with
the result of the usual representation of ChPT at $O(p^4)$.
   In this case the vertex at $q^2=0$ is independent of $p_f^2$ and $p_i^2$,
$\Gamma^{\mu,irr}_R(p_f,p_i)=(p_f+p_i)^\mu$.
   Furthermore, the renormalized propagator is simply given by the free
propagator.

   We shall now address the question how the parameter $\beta_1$ enters
the Compton scattering amplitude \cite{Scherer2}.
   For that purpose we subtract the ordinary calculation of the pole terms
using free vertices from the corresponding calculation with off--shell
vertices and interpret the result as being due to off--shell effects.
   Similar methods have been the basis of investigating the influence of
off--shell form functions in various reactions involving the nucleon, such
as proton--proton bremsstrahlung \cite{Nyman}, electron--nucleus scattering
\cite{Naus}, or virtual Compton scattering \cite{Brand}.
   For Compton scattering from a pion the result is found to
be,\footnote{Of course, using Coulomb gauge
$\epsilon^{\mu}=(0,\vec{\epsilon})$, $\epsilon'^{\mu}=(0,\vec{\epsilon'})$,
and performing the calculation in the lab frame ($p_i^{\mu}=(M_{\pi},0)$),
the additional contribution vanishes, since $p_i\cdot\epsilon=
p_i\cdot\epsilon'=0$.
   However, this is a gauge--dependent statement and thus not true for a
general gauge.}
\begin{eqnarray}
\label{chpt}
\Delta M_P&=&\underbrace{M_P(\beta_1\neq 0)}_{\mbox{incl.\ off--shell
effects}}-\underbrace{M_P(\beta_1=0)}_{\mbox{ordinary calc.}}\nonumber\\
&=&-ie^2 \frac{64 \beta_1}{F^2_{\pi}}
\left(
\underbrace{p_f\cdot\epsilon'\,p_i\cdot\epsilon}_{\mbox{s channel}}
+
\underbrace{p_f\cdot\epsilon\, p_i\cdot\epsilon'}_{\mbox{u channel}}
\right).
\end{eqnarray}
   We shall now demonstrate that Eq.\ (\ref{chpt}) cannot be used for a
unique extraction of the form functions from experimental data.
   In order to see this we have to realize that Eq.\ (\ref{chpt}) is not yet
the complete modification of the total amplitude.
   The reason is that the very same term in the Lagrangian which contributes
to the off--shell electromagnetic vertex also generates a two--photon
contact interaction.
   This can be seen by inserting the appropriate covariant derivative into
Eq.\ (\ref{loffshell}) and by selecting those terms which contain two powers of
the pion field as well as two powers of the electromagnetic field.
   From the first term of Eq.\ (\ref{loffshell}) one obtains the following
$\gamma\gamma\pi\pi$ interaction term
\begin{eqnarray}
\label{loff1}
{\Delta\cal L}_{\gamma\gamma\pi\pi}&=&\frac{16\beta_1 e^2}{F^2_{\pi}}
\left(-A^2[\pi^-(\Box+M^2_{\pi})\pi^+
+\pi^+(\Box+M^2_{\pi})\pi^-]\right.\nonumber\\
&&\left.+(\partial\cdot A+2A\cdot\partial)\pi^+
(\partial\cdot A+2A\cdot\partial)\pi^-\right).
\end{eqnarray}
   For real photons Eq. (\ref{loff1}) translates into a contact contribution
of the form
\begin{equation}
\label{f1}
\Delta {\cal M}_{\gamma\gamma\pi\pi}=ie^2\frac{64\beta_1}{F^2_\pi}
(p_f\cdot\epsilon'\,p_i\cdot
\epsilon+p_f\cdot\epsilon\, p_i\cdot\epsilon'),
\end{equation}
which precisely cancels the contribution of Eq.\ (\ref{chpt}).
   At first sight the second term of Eq.\ (\ref{loffshell}) also seems to
generate a contribution to the Compton scattering amplitude.
   However, after wave function renormalization this term drops out (see
Ref.\ \cite{Scherer2} for details).
   We emphasize that all the cancellations happen only when one consistently
calculates off--shell form functions, propagators and contact terms, and
properly takes renormalization into account.
   Thus the Lagrangian of Eq.\ (\ref{lagup}) which represents an equivalent
form to the standard Lagrangian of ChPT yields the same Compton scattering
amplitude while, at the same time, it generates different off--shell
form functions.
   Clearly, this illustrates why there is no unambiguous way of extracting the
off--shell behaviour of form functions from on--shell matrix elements.
   The ultimate reason is that the form functions of Eq.\ (\ref{par})
are not only model dependent but also representation dependent, i.e.,
two representations of the same theory result in the same observables
but different form functions.

   Finally, let us discuss the above derivation within a somewhat different
approach which does not make use of a calculation within a specific model
or theory.
   Such a discussion also serves to demonstrate that our interpretation is
independent of the fact that we made use of ChPT at $O(p^4)$.
   For that purpose we follow Gell--Mann and Goldberger in their derivation
of the low--energy theorem for Compton scattering \cite{Gell-Mann}, and split
the most general invariant amplitude of
$\gamma (\epsilon,k)+\pi^+(p_i)\rightarrow \gamma'(\epsilon',k')+\pi^+(p_f)$
into two classes $A$ and $B$.
   Class $A$ consists of the most general pole terms and class $B$ contains
the rest,
\begin{equation}
\label{mab}
{\cal M}=\epsilon'_\nu M^{\nu\mu}\epsilon_\mu={\cal M}_A+{\cal M}_B.
\end{equation}
   The original motivation in Ref.\ \cite{Gell-Mann} for such a separation was
to isolate those terms of ${\cal M}$ which have a singular behaviour in the
limit $k,k'\to 0$.
   We write class $A$ in terms of the most general expressions for the
irreducible, renormalized vertices and the renormalized propagator,
\begin{equation}
\label{ma}
M^{\nu\mu}_A=-ie^2 \Gamma^\nu(p_f,p_f+k')\Delta_R (p_i+k)
\Gamma^\mu (p_i+k,p_i)
+ (k\leftrightarrow -k',\mu\leftrightarrow\nu),
\end{equation}
where we made use of crossing symmetry.
   For sufficiently low energies class $B$ can be expanded in terms of the
relevant kinematical variables,
\begin{equation}
\label{mb}
M^{\nu\mu}_B=a^{\nu\mu}(P)+b^{\nu\mu\rho}(P)k_\rho+
c^{\nu\mu\rho}(P)k'_\rho+\dots,
\end{equation}
where $P=p_i+p_f$.
   Furthermore, in class $A$ we expand the vertices and the renormalized
propagator of the pion around their respective on--shell points, $p^2=M^2$.
   We obtain for the propagator
\begin{equation}
\label{exprop}
\Delta^{-1}_R(p^2)=p^2-M^2-\Sigma(p^2)=(p^2-M^2)
(1-\frac{p^2-M^2}{2}\Sigma''(M^2)+\dots),
\end{equation}
where we made use of the standard normalization conditions
$\Sigma(M^2)=\Sigma'(M^2)=0$.
   The expansion of, e.g., the vertex describing the absorbtion of the initial
photon in the s channel reads
\begin{eqnarray}
\label{expvert}
\Gamma^\mu(p_i+k,p_i)&=&(P^\mu+k'^\mu)F(0,M^2+(s-M^2),M^2)\nonumber\\
&=&(P^\mu+k'^\mu)(1+(s-M^2)\partial_2F(0,M^2,M^2)+\dots),
\end{eqnarray}
where we made use of $k^2=0$, and where $\partial_2$ refers to partial
differentiation with respect to the second argument.
   Inserting the result of Eqs.\ (\ref{exprop}) and (\ref{expvert}) into
Eq.\ (\ref{ma}) we obtain for the s channel
\begin{eqnarray}
\label{masexp}
M^{\nu\mu}_s&=&-ie^2 (P^\nu+k^\nu)(1+(s-M^2)\partial_3 F(0,M^2,M^2) +\dots)
\nonumber\\
&\times&\frac{1}{s-M^2}(1+\frac{s-M^2}{2}\Sigma''(M^2)+\dots)\nonumber\\
&\times&(P^\mu+k'^\mu)(1+(s-M^2)\partial_2 F(0,M^2,M^2) +\dots)\nonumber\\
&=&-ie^2\frac{(P^\nu+k^\nu)(P^\mu+k'^\mu)}{s-M^2}+O((s-M^2)^0)\nonumber\\
&=&\mbox{``free'' s channel + analytical terms},
\end{eqnarray}
and an analogous term for the u channel.
   In Eq.\ (\ref{masexp}) ``free'' s channel refers to a calculation with
on--shell vertices.
   From Eq.\ (\ref{masexp}) we immediately see that off--shell effects
resulting from either the form functions or the renormalized propagator are
of the same order as analytical contributions from class $B$.
   In the total amplitude off--shell contributions from class $A$ cannot
uniquely be separated from class $B$ contributions.
   In the language of field transformations this means that contributions
to $\cal M$ can be shifted between different diagrams leaving the total result
invariant.

\section{Conclusions}
   We discussed the electromagnetic vertex of an off--shell pion.
   The most general form of this vertex is constrained by symmetry principles
such as Lorentz covariance, gauge invariance, time--reversal and
charge--conjugation symmetry (see items 1) -- 4) of Sec.\ II).
   In order to illustrate how such an off--shell vertex enters a covariant
calculation of a scattering amplitude, we considered the specific example of
Compton scattering from a pion.
   The calculation was performed within the framework of chiral perturbation
theory at $O(p^4)$.
   We generated an infinite number of equivalent representations of the chiral
Lagrangian by making use of the concept of field transformations.
   This approach allowed us to compare the results of different representations
of the same microscopic theory.

   It was demonstrated that different but physically equivalent representations
generate different off--shell electromagnetic vertices.
   On the other hand, all representations result in the same Compton scattering
amplitude.
   This is a consequence of the equivalence theorem.
   As a result of our specific example we conclude that even in the framework
of the {\em same} microscopic theory, given in different representations,
it is not possible to uniquely extract the contributions to the scattering
amplitude which result from off--shell effects in the pole terms.

   In the language of Gell--Mann and Goldberger, by a change of representation,
contributions can be shifted from class $A$ to class $B$ within the {\em same}
theory.
   We can also express this differently; what appears to be an off--shell
effect in one representation results, for example, from a contact interaction
in another representation.
    In this sense, off--shell effects are not only model dependent, i.e.,
different models generate different off--shell form functions, but they
are also representation dependent which means that even different
representations of the same theory generate different off--shell form
functions.
   This has to be contrasted with on--shell S--matrix elements which,
in general, will be different for different models (model dependent),
but always the same for different representations of the same model
(representation independent).

    We have seen that the most general result for the Compton scattering
amplitude up to $O(p^4)$ can be obtained in a representation ($\beta_1=0$)
with no off--shell effects at all in the electromagnetic vertex for real
photons.
   This is a special feature of the momentum expansion of chiral perturbation
theory up to $O(p^4)$, and one should not generalize this observation to
higher orders in the momentum expansion.
Higher--loop diagrams may, in general, yield off--shell contributions which
cannot be transformed away.

In conclusion, the freedom of performing field transformations allows to
shift contributions between different building blocks in different
representations of the same theory, while the on--shell S--matrix remains the
same.
In general, quantum field theoretical models will yield off--shell vertices,
however, they are not unique.
In particular, they are not only model dependent but also representation
dependent.

\section{Acknowledgements}
   I am very grateful to Harold W.\ Fearing for a fruitful collaboration on
the topic of this talk.
   Furthermore, I would like to thank Justus H.\ Koch for arousing my interest
in this subject.

\end{document}